\documentclass{article}
\usepackage{fullpage}
\usepackage{hyperref}
\usepackage{booktabs}
\usepackage{multirow}
\usepackage{textcomp}

\usepackage{soul}
\usepackage[usenames, dvipsnames]{xcolor}

\author{Marcel Keller, Ke Sun \\ CSIRO's Data61}
\title{A Note on Our Submission to \\ Track 4 of iDASH 2019}

\begin{document}
\maketitle

\begin{abstract}
  iDASH is a competition soliciting implementations of cryptographic
  schemes of interest in the context of biology. In 2019, one track
  asked for multi-party computation implementations of training of a
  machine learning model suitable for two datasets from cancer research.
  In this note, we describe our solution submitted to the
  competition. We found that the training can be run on three AWS
  \texttt{c5.9xlarge} instances in less then one minute using MPC
  tolerating one semi-honest corruption, and less than ten seconds at
  a slightly lower accuracy.
\end{abstract}

\section{Introduction}

In this section, we summarize the
task.\footnote{\url{http://www.humangenomeprivacy.org/2019/competition-tasks.html}}
Participants were invited to implement machine learning (ML) training in
multi-party computation (MPC). MPC allows a set of parties to
jointly compute on data hold among each other without revealing anything other
than the result of the computation. In the current context this means
that a set of healthcare providers holding measurements about cancer
patients and healthy individuals can jointly compute a ML model
detecting cancer without sharing the measurements.

The training algorithm must be suitable to the GSE2034 \cite{gse2034}
and BC-TCGA \cite{BC-TCGA} breast cancer datasets. The organizers provided a subset of both. For GSE2034, this
contains 142 positive (recurrence tumor)
and 83 negative (no recurrence normal) samples,
each with 12,634 features. On the other hand, the subset of BC-TCGA
contains 422 positive (breast cancer tissue)
and 48 negative (normal tissue) samples 
of 17,814 features each. The organizers
provided a reference model, and the submissions were expected to perform similarly.

In terms of security, the competition asked for three-party computation
with one semi-honest corruption. This security model has received
widespread attention because it does not require relatively
expensive cryptographic primitives such as oblivious transfer or
homomorphic encryption. Instead, so-called replicated secret sharing
suffices, where every party holds two out of three random shares which
sum up to a secret value~\cite{CCS:AFLNO16}. Furthermore, semi-honest
security requires that even corrupted parties follow the protocol,
which allows the creation of optimized protocols for specific
purposes, for example probabilistic truncation
\cite{cryptoeprint:2019:131}.

For the evaluation, the organizers asked for Docker containers that
would be run on three hosts in a local cluster. The submissions were
required to finish within 24 hours, and they were ranked on accuracy,
performance, and communication.


\section{The Model}

\renewcommand{\P}{\mathrm{P}}
\newcommand{\N}{\mathrm{N}}
\newcommand{\TP}{\mathrm{TP}}
\newcommand{\TN}{\mathrm{TN}}
\newcommand{\FP}{\mathrm{FP}}
\newcommand{\FN}{\mathrm{FN}}

We evaluated the performance of the following baseline models using plaintext computation
\begin{description}
\item[Logistic] Logistic regression based on mini-batch stochastic gradient
descent (SGD) with constant learning rate and momentum.
The model is trained for a fixed number of 100 epochs. In each mini-batch of 16 samples,
we re-balance the positive and negative classes by re-sampling the same number of
samples from these two classes;
\item[MLP] A multilayer perceptron with two hidden layers of 256+64 ReLU~\cite{relu} neurons.
We apply dropout~\cite{dropout} with probability 0.5 to the hidden layers to avoid over-fitting. The
model is trained by SGD for 100 epochs using the same mini-batches as Logistic;
\item[Linear SVM] A linear SVM classifier with $L_2$ regularization and typical settings.
We use scikit-learn's~\cite{sklearn} implementation with a stopping tolerance of
$10^{-3}$ and other default settings;
\item[Random Forest] A random forest classier~\cite{tin95} with 100 trees
based on scikit-learn's~\cite{sklearn} implementation;
\item[Reference] The reference model provided by the iDASH committee, which is a
deep 1D convolutional network composed of Residual blocks~\cite{resnet}.
We train the model using the Adam optimizer~\cite{adam} with a fixed learning rate of 
$10^{-4}$ for 30 epochs.
\end{description}
In these models, both Logistic and Linear SVM correspond to a single-neuron model.
The main difference between these two methods is the loss function:
Logistic minimizes the cross-entropy between the target label and the prediction,
while Linear SVM minimizes the squared Hinge loss.
The detailed hyper-parameter configurations are omitted for brevity. We tried to
achieve a representative accuracy score for each method. Further tuning these
methods can give marginal improvement.  We tried to compare all methods using
similar settings such as how the mini-batches are constructed and how the
performance is evaluated.

Denote $\TP$ (true positive), $\FN$ (false negative) to be the
population of the ground truth positive class; denote $\TN$ (true negative), and
$\FP$ (false positive) to be the negative class. Then the $F_1$ score
of these two classes are given by the harmonic mean of the precision and recall,
that is,
\begin{equation}
F^{\P}_1 = \frac{2}{\frac{1}{\frac{\TP}{\TP+\FP}} + \frac{1}{\frac{\TP}{\TP+\FN}}} = \frac{2\TP}{2\TP + \FP + \FN},
\quad
F^{\N}_1 = \frac{2\TN}{2\TN + \FP + \FN}.
\end{equation}
Then we evaluate the classification performance based on the weighted $F_1$ score
\begin{equation}
F_1 = \frac{\TP+\FN}{\TP+\TN+\FP+\FN} F^{\P}_1 + \frac{\TN+\FP}{\TP+\TN+\FP+\FN} F^{\N}_1,
\end{equation}
which is a population-weighted mean of $F^{\P}_1$ and $F^{\N}_1$.

See Table~\ref{table:models} for our cross-validation accuracy scores and timing results.
Observe that MLP does not improve over Logistic, because the training data is
linearly separable and prone to over-fitting.
The poor performance of the reference model, despite its expensive training
cost, is due to over-fitting. One may improve further its performance with 
a carefully designed optimization procedure, e.g., based on dynamic learning rates.
However, the benefit is quite limited in light of the high complexity to implement it.

Eventually, we decided to go ahead with logistic regression because
its accuracy comes close the
other models while its simplicity promised a more efficient
implementation.
In the next section, we will show that our MPC implementation
slightly surpasses the plaintext implementation of logistic regression and
comes even closer to the linear SVM classifier in terms of accuracy.

\begin{table}
  \centering
  \begin{tabular}{@{}r|cc|cc}
    \toprule
                & \multicolumn{2}{c|}{GSE2034} & \multicolumn{2}{c}{BC-TCGA} \\
     Model      & $F_1$ & Time (sec) & $F_1$ & Time (sec) \\
    \midrule
    Logistic           & $0.666\pm0.062$ & 3  & $0.995\pm0.006$ & 8 \\
    MLP                & $0.664\pm0.066$ & 35 & $0.994\pm0.007$ & 39 \\
    Linear SVM         & $0.677\pm0.058$ & 1 & $0.996\pm0.006$ & 0.3\\
    Random Forest      & $0.592\pm0.055$ & 0.6 & $0.988\pm0.011$ & 0.8\\
    Reference          & $0.650\pm0.075$ & 68$^\star$ & $0.987\pm0.008$ & $139^\star$ \\
    \bottomrule
  \end{tabular}
  \caption{Weighted $F_1$ score and computational time in seconds with plaintext computation.
All reported number are averages over 20 runs of five-fold cross validation (100 folds in total).
The reported time in seconds is measured on an Intel~Core~i5-7300U CPU.
The upper-script ``$^\star$'' means that the timing is performed instead on a NVidia Tesla P100
because the experiments could not finish in reasonable time.}
  \label{table:models}
\end{table}


\section{Our Implementation}

We implemented our solution in MP-SPDZ \cite{mp-spdz}. It features
fixed-point computation, that is, the fractional number $x$ is
represented as an integer near $x \cdot 2^k$ for some $k$. Addition and
subtraction are straight-forward due to linearity while
multiplication is implemented as integer multiplication followed by
truncation. This truncation can either mean rounding to the nearest
integer or the more efficient probabilistic truncation where rounding
down is the more likely the closer the input number is to the
floor. See Catrina and Saxena~\cite{FC:CatSax10} for more details.

Standard logistic regression uses the sigmoid function based on the
exponential, and computing the loss requires the logarithm
function. Our implementation of these is based on the code provided in
SCALE-MAMBA \cite{scale-mamba} by Aly and Smart \cite{ACNS:AlySma19}.

A particular helpful feature of MP-SPDZ in some protocols including
the one used here is the implementation of dot products of fixed-point
numbers with constant communication. See Dalskov et
al. \cite{cryptoeprint:2019:131} for more details. Due to this
optimization, we decided not to use mini-batches but the whole training
set at once for simplicity. 
The benefit of the optimization increases with the size of the batch.
Our implementation can be straightforwardly generalized to mini-batch training.

Given the decisions above, we ran our implementation with a few
parameters on AWS \texttt{c5.9xlarge}, namely the precision of
fixed-point truncation and the duration. For the former, there is a
choice of probabilistic and exact truncation.
Considering the latter, we saw that
it takes about 100 epochs for the loss to get close to zero without
using mini-batches. Therefore, we ran our implementation either for 
100 or for 200 epochs, or until the loss was below $10^{-4}$.

Table \ref{table:results} shows the five-fold cross-validation accuracy
and running time
for each combination of parameters and dataset. Each values is
averaged over 100 runs. Note that the running times are taken from the
same run as the accuracies and therefore use only 80\% of the
respective dataset.

\begin{table}
  \centering
  \begin{tabular}{@{}lllcc@{}}
    \toprule
    Dataset & Duration & Truncation
    & $F_1$
    & Time (sec) \\
    \midrule
    \multirow{6}{*}{GSE2034} & \multirow{2}{*}{100} & Probabilistic
    & $0.670 \pm 0.070$
    & 8 \\
            && Exact
                        & $0.666 \pm 0.068$
    & 20 \\
            & \multirow{2}{*}{200} & Probabilistic
                        & $0.674 \pm 0.063$
    & 14 \\ 
            && Exact
                        & $0.670 \pm 0.066$
    & 38 \\
            & \multirow{2}{*}{Variable} & Probabilistic
                         & $0.657 \pm 0.091$
    & 6 \\ 
            && Exact
                        & $0.650 \pm 0.091$
    & 22 \\
    \midrule
    \multirow{6}{*}{BC-TCGA} & \multirow{2}{*}{100} & Probabilistic
                 & $0.994 \pm 0.007$
    & 21 \\
            && Exact
                        & $0.994 \pm 0.008$
                        & 43 \\
            & \multirow{2}{*}{200} & Probabilistic
    & $0.994 \pm 0.007$
    & 38 \\ 
            && Exact
                        & $0.994 \pm 0.009$
    & 77 \\
            & \multirow{2}{*}{Variable} & Probabilistic
                         & $0.995 \pm 0.010$
    & 8 \\ 
            && Exact
                        & $0.994 \pm 0.011$
    & 17 \\
    \bottomrule
  \end{tabular}
  \caption{Five-fold cross-validation accuracy and running times of our
    implementation}
  \label{table:results}
\end{table}

The implementation used for these timings is different to the
submitted version in two points: The submitted version would only use
a single thread because the evaluation criterion was changed from a
single host to several shortly before the deadline. Furthermore, the
improved version uses the dot product optimization more
consequently. We have seen that this reduces the running time by about one
third.

Given the result, one would choose 200 epochs with probabilistic
rounding for the best accuracy and variable duration with
probabilistic rounding for faster inference at the cost of accuracy.
However, we did not have
time to run these evaluations before the deadline. From the limited
information we had at the time, we decided to submit the training with
200 epochs and exact rounding.
We achieved slightly better accuracy than logistic regression with plaintext
computation, because a limited precision helps improve generalization on these two datasets.


\bibliographystyle{abbrv}
\bibliography{note,cryptobib/abbrev3,cryptobib/crypto}

\begin{thebibliography}{10}

\bibitem{ACNS:AlySma19}
A.~Aly and N.~P. Smart.
\newblock Benchmarking privacy preserving scientific operations.
\newblock In R.~H. Deng, V.~{Gauthier-Uma{\~n}a}, M.~Ochoa, and M.~Yung,
  editors, {\em ACNS 19}, volume 11464 of {\em {LNCS}}, pages 509--529.
  Springer, Heidelberg, June 2019.

\bibitem{CCS:AFLNO16}
T.~Araki, J.~Furukawa, Y.~Lindell, A.~Nof, and K.~Ohara.
\newblock High-throughput semi-honest secure three-party computation with an
  honest majority.
\newblock In E.~R. Weippl, S.~Katzenbeisser, C.~Kruegel, A.~C. Myers, and
  S.~Halevi, editors, {\em ACM CCS 2016}, pages 805--817. {ACM} Press, Oct.
  2016.

\bibitem{FC:CatSax10}
O.~Catrina and A.~Saxena.
\newblock Secure computation with fixed-point numbers.
\newblock In R.~Sion, editor, {\em FC 2010}, volume 6052 of {\em {LNCS}}, pages
  35--50. Springer, Heidelberg, Jan. 2010.

\bibitem{scale-mamba}
{COSIC, KU Leuven}.
\newblock {SCALE-MAMBA}.
\newblock \url{https://github.com/KULeuven-COSIC/SCALE-MAMBA}, 2019.

\bibitem{cryptoeprint:2019:131}
A.~Dalskov, D.~Escudero, and M.~Keller.
\newblock Secure evaluation of quantized neural networks.
\newblock Cryptology ePrint Archive, Report 2019/131, 2019.
\newblock \url{https://eprint.iacr.org/2019/131}.

\bibitem{mp-spdz}
{Data61}.
\newblock {MP-SPDZ}.
\newblock \url{https://github.com/data61/MP-SPDZ}, 2019.

\bibitem{relu}
X.~Glorot, A.~Bordes, and Y.~Bengio.
\newblock Deep sparse rectifier neural networks.
\newblock In {\em Proceedings of the Fourteenth International Conference on
  Artificial Intelligence and Statistics}, volume~15 of {\em Proceedings of
  Machine Learning Research}, pages 315--323. PMLR, 2011.

\bibitem{resnet}
K.~{He}, X.~{Zhang}, S.~{Ren}, and J.~{Sun}.
\newblock Deep residual learning for image recognition.
\newblock In {\em CVPR}, pages 770--778, 2016.

\bibitem{tin95}
T.~K. Ho.
\newblock Random decision forests.
\newblock In {\em International Conference on Document Analysis and
  Recognition}, pages 278--282, 1995.

\bibitem{adam}
D.~P. Kingma and J.~Ba.
\newblock Adam: A method for stochastic optimization.
\newblock In {\em ICLR}, 2015.
\newblock \url{https://arxiv.org/abs/1412.6980}.

\bibitem{BC-TCGA}
C.~G.~A. Network et~al.
\newblock Comprehensive molecular portraits of human breast tumours.
\newblock {\em Nature}, 490(7418):61, 2012.

\bibitem{sklearn}
F.~Pedregosa, G.~Varoquaux, A.~Gramfort, V.~Michel, B.~Thirion, O.~Grisel,
  M.~Blondel, P.~Prettenhofer, R.~Weiss, V.~Dubourg, J.~Vanderplas, A.~Passos,
  D.~Cournapeau, M.~Brucher, M.~Perrot, and E.~Duchesnay.
\newblock Scikit-learn: Machine learning in {P}ython.
\newblock {\em Journal of Machine Learning Research}, 12:2825--2830, 2011.

\bibitem{dropout}
N.~Srivastava, G.~Hinton, A.~Krizhevsky, I.~Sutskever, and R.~Salakhutdinov.
\newblock Dropout: A simple way to prevent neural networks from overfitting.
\newblock {\em Journal of Machine Learning Research}, 15:1929--1958, 2014.

\bibitem{gse2034}
Y.~Wang, J.~G. Klijn, Y.~Zhang, A.~M. Sieuwerts, M.~P. Look, F.~Yang,
  D.~Talantov, M.~Timmermans, M.~E. Meijer-van Gelder, J.~Yu, et~al.
\newblock Gene-expression profiles to predict distant metastasis of
  lymph-node-negative primary breast cancer.
\newblock {\em The Lancet}, 365(9460):671--679, 2005.

\end{thebibliography}

\end{document}